\documentclass[twocolumn]{jpsj3} 
\usepackage{bm}

\title{Coherence Effect of Sign-Reversing $s_\pm$-wave Cooper Pair State in Heavily Overdoped LaFeAsO-based Superconductor: $^{75}$As-Nuclear Quadrupole Resonance }

\author{
Hidekazu Mukuda$^{1,3}$\thanks{E-mail address: mukuda@mp.es.osaka-u.ac.jp}, Mariko Nitta$^{1}$, Mitsuharu Yashima$^{1,3}$, Yoshio Kitaoka$^{1}$,\\ Parasharam M. Shirage$^{2}$, Hiroshi Eisaki$^{2,3}$, and Akira Iyo$^{2,3}$ 
}

\inst{
$^{1}$Graduate School of Engineering Science, Osaka University, Toyonaka, Osaka 560-8531 \\
$^{2}$National Institute of Advanced Industrial Science and Technology (AIST), Umezono, Tsukuba 305-8568 \\
$^{3}$JST, TRIP (Transformative Research-Project on Iron Pnictides), Chiyoda, Tokyo 102-0075 
}

\abst{
We report an $^{75}$As-nuclear quadrupole resonance (NQR) study on heavily electron-doped LaFeAsO$_{1-x}$F$_x$(La1111) with $T_c=5$ K. Nuclear spin relaxation rate ($1/T_1$) measurement has revealed that a Hebel-Slichter (HS) peak partially recovers in heavily electron-overdoped regimes where the nesting condition of hole and electron Fermi surfaces(FSs) becomes significantly worse. 
This is in contrast to previous results reported in optimally doped La1111 with $T_c=28$ K where a lack of the HS peak was reported. 
It indicates that the interband scattering between the hole and electron FSs is strongly suppressed by an almost vanishing hole FS through the heavily electron-overdoping. 
Our findings strongly suggest that the sign reversal of the gap functions on the different FSs, that is, $s_\pm$-wave state is realized in La1111 compounds.
We remark that interband scattering on well-nested FSs is essential for stabilizing the $s_\pm$-wave state and enhancing the $T_c$ up to 28 K in LaFeAsO-based superconductors. 
}

\kword{superconductivity, iron-based oxypnictide, LaFeAsO, NQR}

\begin{document}

\maketitle

\date{\today}


The coherence effect of superconductivity(SC), appearing as a Hebel-Slichter(HS) peak in the nuclear spin relaxation rate ($1/T_1$), was one of the crucial experimental proofs for Bardeen-Cooper-Schrieffer (BCS) theory, characterized by conventional $s$-wave Cooper pairs with an isotropic gap\cite{BCS,Tinkham,HS}.
Newly discovered SC in iron (Fe)-based pnictides\cite{Kamihara2008} is a semimetal with a multiband nature derived from the disconnected hole and electron Fermi surfaces (FSs) at $\Gamma$- and $M$-points, respectively\cite{Singh}.
Due to the nesting of these FSs, the parent material LaFeAsO shows a stripe antiferromagnetic (AFM) order with ${\bf Q}=(0,\pi)$ or $(\pi,0)$ \cite{Cruz}, but the substitution of fluorine for oxygen and/or oxygen deficiencies in the LaO layer causes the lattice compression and provides electrons for the system, yielding a novel SC in LaFeAsO$_{1-x}$F$_x$(La1111) with $T_c=26$ K\cite{Kamihara2008}. 
Theoretically, sign reversal in the SC gap functions on these FSs, namely $s_\pm$ wave state, has been proposed as one of the candidates for Fe-pnictide SC\cite{Mazin,Kuroki}.
In the SC state, although the multiple isotropic SC gaps on the FSs have been unraveled experimentally by angle-resolved-photoemission spectroscopy (ARPES) and penetration depth measurements\cite{Ding,Hashimoto}, the HS peak in $1/T_1$ measurements was not observed\cite{Nakai,Graf,MukudaNQR,Kawasaki}.  
To reconcile this inconsistency, it was theoretically pointed out that these experiments can be explained consistently in terms of a sign reversal fully gapped $s_\pm$ wave scenario by assuming strong interband scattering due to the nesting of the disconnected FSs \cite{NMRtheory1,NMRtheory1a,NMRtheory1b,NMRtheory1c,NMRtheory2}. 
It motivates us to investigate the SC state of extremely overdoped regimes because the nesting condition of the FSs becomes significantly worse.    

In this Letter, we report $^{75}$As-NQR-$1/T_1$ measurement on heavily-electron-doped LaFeAsO$_{1-x}$F$_x$ with $T_c=5$ K  that exhibits a small HS peak derived from the coherence effect of Cooper pairs, whereas it was strongly reduced in the optimally doped La1111(OPT) with $T_c=$28 K \cite{Nakai,MukudaNQR}. 
Through the coherence effect on two compounds, we suggest that the sign reversal of the gap functions in the SC state of La1111 system may be stabilized by the interband scattering on the well nested FSs.

\begin{figure}[tbp]
\begin{center}
\includegraphics[width=0.85\linewidth]{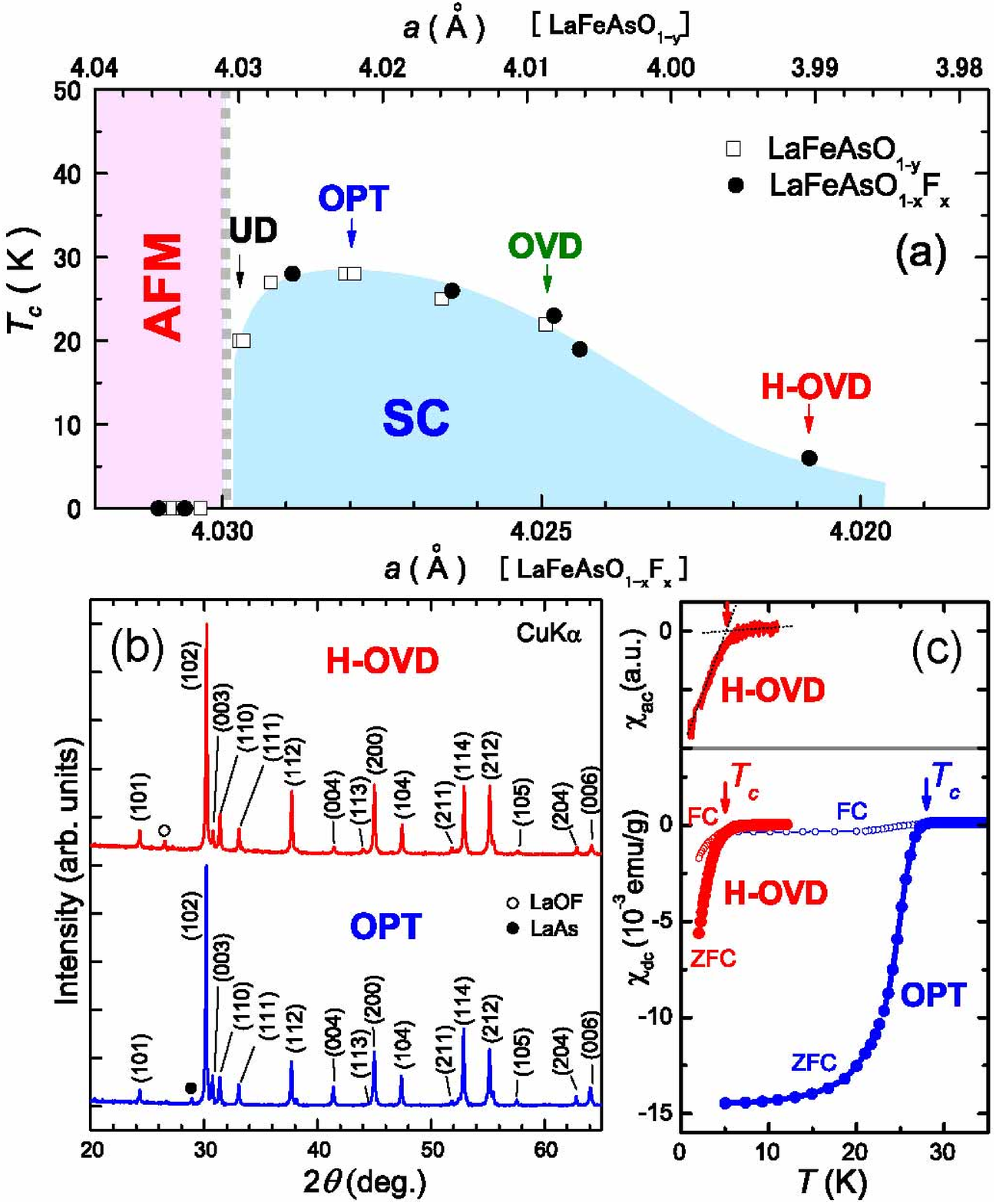}
\end{center}
\caption[]{(color online) (a) $T_c$s of LaFeAsO$_{1-x}$F$_x$ and LaFeAsO$_{1-y}$ as functions of $a$-axis length. (b) XRD patterns of H-OVD and OPT. (c) The $T_c$s are determined by the onset of SC diamagnetism in dc and ac susceptibility measurements. 
}
\label{phasediagram}
\end{figure}

A polycrystalline sample of heavily overdoped LaFeAsO$_{1-x}$F$_x$, denoted as H-OVD, was synthesized by solid state reaction starting with nominal compositions of $x=0.22$. 
Powder x-ray diffraction (XRD) measurement indicates that the sample is almost entirely composed of a single phase, as shown in Fig. \ref{phasediagram}(b). 
$T_c$ was uniquely determined to be 5 K by the onset of SC diamagnetism in both dc and ac susceptibility($\chi_{dc}$ and $\chi_{ac}$) measurements, as shown in Fig. \ref{phasediagram}(c). 
In particular, the $\chi_{ac}$ was measured by using an in situ NQR coil. 
$T_c$s of the H-OVD and the other F-doped and O-deficient La1111 compounds are summarized in Fig.~\ref{phasediagram}(a) as functions of $a$-axis length evaluated at room temperature. 
Actually, the present H-OVD is in a more overdoped region compared with the previous three samples in under-doped (UD), optimally-doped (OPT) and overdoped (OVD) states\cite{Terasaki,MukudaFe2}.
We performed the $^{75}$As-nuclear quadrupole resonance (NQR) experiment at zero external fields on the H-OVD. 

\begin{figure}[tbp]
\begin{center}
\includegraphics[width=0.8\linewidth]{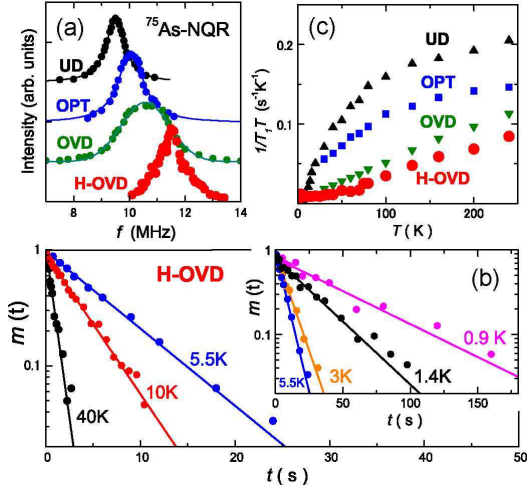}
\end{center}
\caption[]{(color online) 
(a) $^{75}$As-NQR spectrum of H-OVD at 10 K, along with the other La1111 compounds\cite{MukudaNQR,MukudaPhysC}. (b) Nuclear recovery curves $m(t)$ for H-OVD, which is used to determine the $1/T_1$. 
(c) $T$ dependence of $1/T_1T$ in H-OVD, along with that of the others($^{75}$As-NMR data in ref. 21). The high electron doping level of H-OVD is corroborated from the positive shift of NQR frequency and the decrease of $1/T_1T$ value in the normal state. 
}
\label{NQR_recovery}
\end{figure}


The $^{75}$As-NQR spectrum of H-OVD was observed at $^{75}\nu_Q\approx$11.5 MHz, as shown in Fig. \ref{NQR_recovery}(a). When noting the fact that the value of $^{75}\nu_Q$ in La1111 systems continuously increases with increasing doping levels \cite{MukudaNQR,MukudaPhysC,SKitagawa,Lang}, the largest value of $^{75}\nu_Q$ ensures the sufficiently high doping level of H-OVD. The linewidth of the spectrum was as narrow as $\sim$1 MHz, which was comparable to the OPT. 
The nuclear spin-lattice relaxation rate $(1/T_1)$ of $^{75}$As-NQR was measured at $f=$11.5 MHz in a wide temperature($T$) range from 0.8 K to 240 K, and was obtained by fitting with a theoretical recovery curve of $^{75}$As-nuclear magnetization ($I=3/2$), which is expressed by a simple exponential function as $m(t)\equiv(M_0-M(t))/M_0=\exp(-3t/T_{1})$, where $M_0$ and $M(t)$ are the respective nuclear magnetizations for the thermal equilibrium condition and at a time $t$ after the saturation pulse. 
Note that the $1/T_1$ was uniquely determined by a single exponential function of $m(t)$ in both the SC and normal states, as shown in Fig. \ref{NQR_recovery}(b). 
These results ensure that the electronic state of H-OVD is homogeneous throughout the sample even from a microscopic point of view. 
Figure \ref{NQR_recovery}(c) shows the $T$ dependence of $1/T_1T$ in the normal state. 
The $1/T_1T$ of H-OVD remarkably decreases upon cooling, the rate of which is steeper than the previous results on other La1111 systems\cite{MukudaNQR,MukudaFe2}.
This is primarily attributed to a decrease in the density of states (DOS) at the Fermi level($E_F$), namely, the band structure effect derived from the existence of a high DOS just below the $E_F$\cite{Ikeda}. 
Moreover, the value of $1/T_1T$ at 240 K decreases with increasing electron-doping levels, which is reasonably explained by an energy shift in the framework of a rigid-band model. 
It is noteworthy that the $1/T_1T$ of H-OVD stays constant from 50 K to $T_c$=5 K, indicating that the $E_F$ in H-OVD locates in the broad minimum of DOS\cite{Singh}.  
In fact, the ratio of the normal-state DOS at $E_F$ for H-OVD and OPT, $(N_0^{\rm H-OVD}/N_0^{\rm OPT})$, is approximately 0.47, which was evaluated from the relation of $1/T_1T\sim N_0^2$ just above $T_c$.

\begin{figure}[tbp]
\begin{center}
\includegraphics[width=0.8\linewidth]{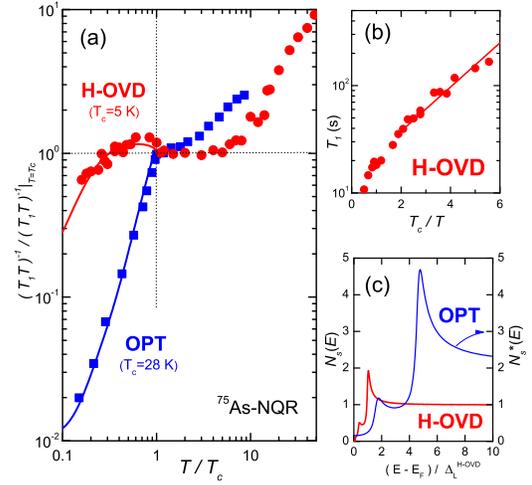}
\end{center}
\caption[]{(color online) (a) $1/T_{1}T$ normalized by the value at $T=T_c$ for H-OVD and OPT as a function of the normalized temperature $T/T_c$.  (b) The plot of $T_1$ vs. $T_c/T$ enables us to evaluate $2\Delta_0/k_BT_c\sim0.95$ roughly from the slope.
Solid curves in (a) for OPT and OVD are simulations based on the two-fully-gapped $s_\pm$-wave model with parameters $2\Delta_L/k_BT_c$ and $\alpha_c$ (See text). 
(c) The DOS used in the simulation\cite{Yashima}.  Note that $N_s(E)$ of OPT is displayed as $N_s(E)/0.47$ by using the ratio of the normal-state DOS of OPT and H-OVD just above $T_c$.  
}
\label{invT1T}
\end{figure}

In the SC state with a momentum dependent SC gap $\Delta(\phi,\theta)$, $1/T_1$ is generally given by
\begin{align*}
\frac{(T_1T)^{-1}}{(T_1T)^{-1}_{T=T_{\rm c}}}=&\frac{2}{k_{\rm B}T_{\rm c}}\int^{\infty}_{0}\left[N_{\rm s}(E)^2+\alpha_c M_{\rm s}(E)^2\right]\\
&\times f(E)\left[1-f(E)\right]dE \hspace{20mm}(1)\\
N_{\rm s}(E)=&\frac{1}{4\pi}\int^{2\pi}_{0}\int^{\pi}_{0}\frac{E}{\sqrt{E^2-|\Delta(\phi,\theta)|^2}}\sin\theta d\theta d\phi\\
M_{\rm s}(E)=&\frac{1}{4\pi}\int^{2\pi}_{0}\int^{\pi}_{0}\frac{\Delta(\phi,\theta)}{\sqrt{E^2-|\Delta(\phi,\theta)|^2}}\sin\theta d\theta d\phi
\end{align*}
where $N_{\rm s}(E)$ and $M_{\rm s}(E)$ are the DOS for quasiparticles and the anomalous DOS originating from the coherence effect of the transition probability in the SC state, respectively. 
In conventional $s$-wave SC, the presence of $M_{\rm s}(E)$ gives rise to the HS peak in $1/T_1T$ just below $T_c$\cite{Tinkham,HS} since it usually has an isotropic gap with the same sign on the all FSs. 
By contrast, in unconventional $d$-wave and/or $p$-wave SC states, the $M_{\rm s}(E)$ term is cancelled out by integrating over the momentum space on the SC gap. 
In the multiband system, the $N_s(E)$ and $M_s(E)$ terms in eq. (1) are represented as $(N_s^h(E)+N_s^e(E))$ and $(M_s^h(E)+M_s^e(E))$, respectively, where the $N_s^h(E)$ and $N_s^e(E)$ are the DOS of the hole and electron FSs, respectively. 
As for the Fe-pnictides, apparent trace of the HS peak has not been reported to date\cite{Nakai,
Graf,MukudaNQR,Kawasaki,Terasaki}, implying that the $M_{\rm s}(E)$ is negligibly small. 
It was theoretically proposed that this result is accounted for on a basis of a nodeless $s_\pm$-wave pairing scenario assuming a sign reversal gap function, $+\Delta_h$ and $-\Delta_e$ on the hole and electron FSs, respectively\cite{NMRtheory1,NMRtheory1a,NMRtheory1b,NMRtheory1c,NMRtheory2}. 
In cases where the $\Delta_h$ and $\Delta_e$ have opposite signs, it is noteworthy that the $2M_s^h(E)M_s^e(E)$ component in $(M_s^h(E)+M_s^e(E))^2$ becomes negative.
In particular, when assuming the well-nested FSs, it is anticipated that the sign-nonconserving interband scattering process ($+\Delta_h\Leftrightarrow-\Delta_e$) may exceed the sign-conserving intraband scattering process ($+\Delta_{h}\Leftrightarrow +\Delta_{h}$ and $-\Delta_{e}\Leftrightarrow -\Delta_{e}$); The former process reduces the $M_s(E)^2$ term through the negative contribution of the $M_s^h(E)M_s^e(E)$, whereas the latter process does not. 
Here, to deal with such convoluted intraband and interband contributions in the nuclear spin relaxation process, we introduce the coefficient $\alpha_c$ in eq. (1) phenomenologically, which takes a value $\alpha_c\le 1$ depending on the weight of the interband contribution. 

Figure \ref{invT1T}(a) shows $1/T_1T$ for H-OVD and OPT, normalized by the value at their $T_c$s as a function of normalized temperature $T/T_c$. 
In the case of OPT presented in the previous paper\cite{MukudaNQR,Yashima,NMRtheory2}, $1/T_1$ was actually reproduced by the two-fully gapped $s_\pm$-wave model with parameters $2\Delta_L/k_BT_c\sim4.4$, $\Delta_S/\Delta_L=0.35$, and a smearing factor of $\eta\sim0.07\Delta_L$ derived from the energy broadening of quasiparticles, then assuming $\alpha_c \sim 0$.
Here, the $\Delta_L$ and $\Delta_S$ are the larger and smaller gaps in the model, respectively. 
The $N_s(E)$ used in the simulation is shown in Fig. \ref{invT1T}(c).
In contrast, the $1/T_1T$ of H-OVD shows  a small HS peak below $T_c$, indicating the non-negligible contribution of the $M_{\rm s}(E)$ term, that is, $\alpha_c \ne 0$. 
In order to evaluate the SC gap, we plot $T_1$ against $T_c/T$, as shown in Fig. \ref{invT1T}(b). 
The slope of this plot enables us to estimate $2\Delta_0/k_BT_c\sim0.95$ as a possible value of $\Delta_L$ when we assume $1/T_1\propto \exp(-\Delta_0/k_BT)$ at $T$ being sufficiently lower than $T_c$.
As shown by the solid line in Fig. \ref{invT1T}(a), the experimental result for H-OVD is well reproduced by the same model as in OPT with  different parameters, $\Delta_L\approx \Delta_0$ and $\alpha_c\sim 1/3$, and with similar parameters, $\Delta_S/\Delta_L\sim 0.35$ and $\eta\sim0.07\Delta_0$.
In this analysis, $2\Delta_L/k_BT_c\sim0.95$ was remarkably smaller than 3.52 in BCS theory, indicating that a very weak coupling SC state is realized in the H-OVD. 
This is in contrast to the $2\Delta_L/k_BT_c\sim6.9$ evaluated in Y-substituted La$_{0.8}$Y$_{0.2}$1111($T_c=34$ K), in which the strong coupling state takes place to increase $T_c$ up to 34 K in LaFeAsO-based compounds \cite{Yamashita2}.
It reasonably coincides with the fact that the H-OVD locates on the verge of an SC phase.  

\begin{figure}[tbp]
\begin{center}
\includegraphics[width=0.85\linewidth]{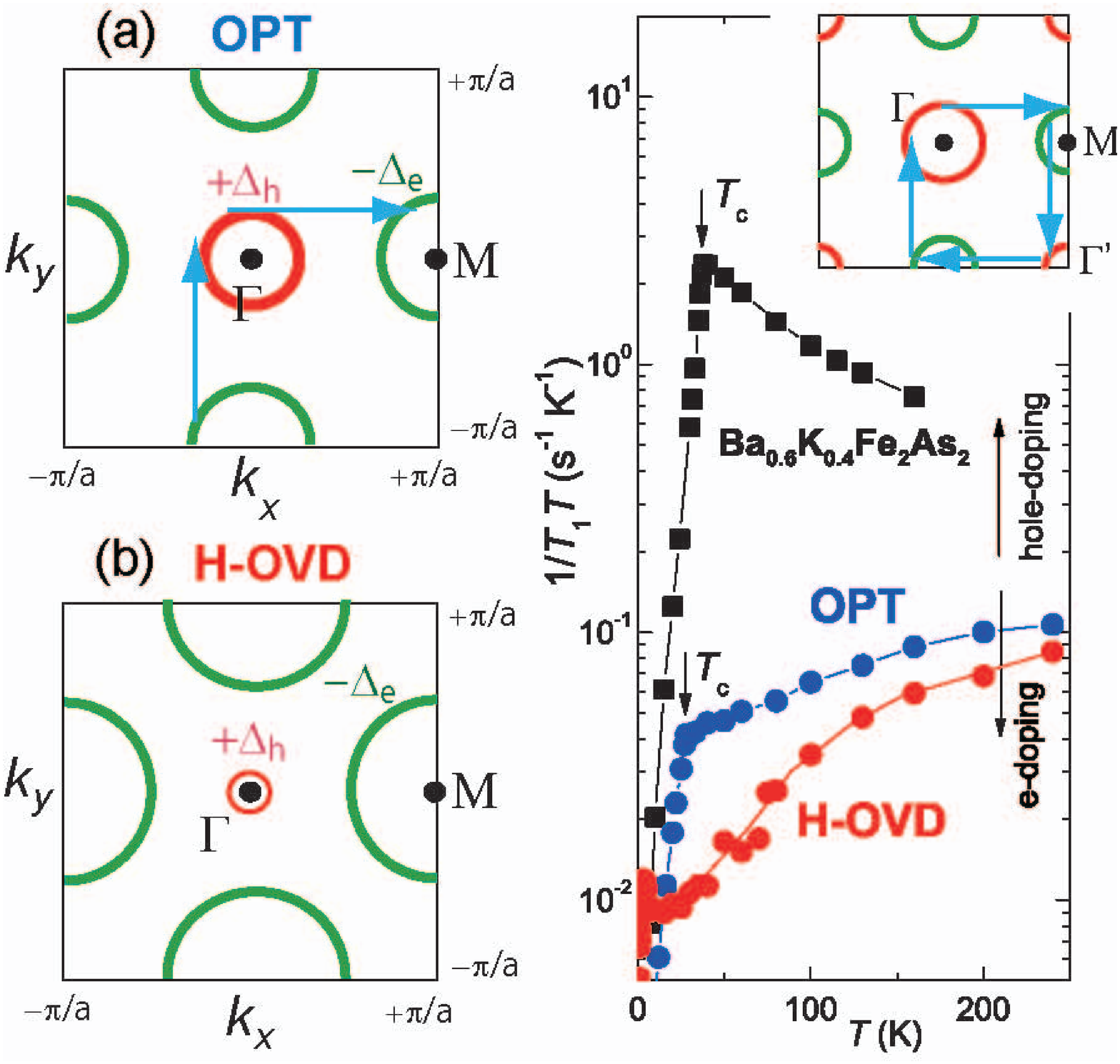}
\end{center}
\caption[]{(color online) Schematics of FS topologies anticipated for (a) OPT($T_c=28$ K) and (b) H-OVD($T_c=5$ K). In OPT, the disconnected FSs derived from hole and electron bands at respective $\Gamma$- and $M$-points are well connected with nesting vector ${\bf Q}=(0,\pi)$ or $(\pi,0)$ \cite{Singh,Kuroki}. The further increase in electron-doping level for H-OVD may make this nesting condition worse, resulting in the dramatic suppression of the interband scattering. 
(c) $T$ dependence of $1/T_1T$ from electron-doped La1111(OPT and H-OVD) to hole-doped Ba$_{0.6}$K$_{0.4}$Fe$_{2}$As$_{2}$($T_c=38$ K)\cite{Yashima}. The inset shows the FS topology theoretically anticipated when the holes are doped and/or the pnictgen height is relatively high\cite{Kuroki2,Ikeda2}. 
}
\label{FS}
\end{figure}

We highlight that the experimental data of $1/T_1T$ in H-OVD cannot be accounted for without assigning $\alpha_c\sim 1/3$ in the analysis. 
This marked difference in $\alpha_c$ between H-OVD and OPT originates from Fermi surface topologies undergoing a chemical potential shift in the rigid band picture through the electron doping, as schematically-illustrated in Figs. \ref{FS}(a) and \ref{FS}(b). 
In the OPT, the respective hole and electron FSs at $\Gamma$- and $M$- points are well nested by the wave vector ${\bf Q}=(\pi, 0)$ and $(0, \pi)$ in the unfolded Brillouin zone\cite{Singh,Kuroki}. 
By contrast, it is expected that the hole FS becomes considerably smaller as a result of the excess electron-doping in H-OVD, in which the nesting condition of these FSs becomes significantly worse. 
The $1/T_1$ of OPT is dominated by a sign-nonconserving interband scattering process ($+\Delta_h\Leftrightarrow-\Delta_e$) that is sufficiently larger than the sign-conserving intraband scattering process ($\pm\Delta_{h,e}\Leftrightarrow\pm\Delta_{h,e}$). 
In H-OVD, however, the interband and intraband contributions become comparable since the interband scattering is significantly suppressed. 
It may cause a partial recovery of the coherence effect of the Cooper pair.  
Actually, the ARPES on the heavily electron-overdoped BaFe$_{1.7}$Co$_{0.3}$As$_{2}$ suggested that the disappearance of the hole FS brings about the absence of $T_c$\cite{Sekiba}. 
The reason why the $T_c$ of H-OVD decreases significantly to 5 K can be attributed to the fact that the interband scattering between these FSs is markedly suppressed. 
In this context, it is the interband scattering which increases $T_c$ up to 28 K in OPT where the nesting condition of the FSs becomes better. 
The results give a strong experimental support for the sign reversing $s_\pm$-wave state in La1111 superconductors. 
This is consistent with recent scanning tunneling microscopy on Fe(Se,Te) that revealed the sign reversal of the SC gap function by measuring the magnetic-field dependence of quasi-particle scattering amplitudes\cite{Hanaguri}. 

Finally, we comment on a relation between the FS topologies and the normal-state properties through the $1/T_1T$ results(see Fig. \ref{FS}(c)). 
The AFM spin fluctuations (AFM-SF) at ${\bf Q}=(\pi, 0)$ and $(0, \pi)$ are pronounced in hole-doped Ba$_{0.6}$K$_{0.4}$Fe$_{2}$As$_{2}$($T_c=38$ K)\cite{Yashima}, but they are not in OPT and H-OVD in La1111 compounds\cite{Terasaki,Yamashita2}. 
The theory suggested that this difference in AFM-SF is ascribed to a possible evolution of FS topologies in these compounds; as schematically-illustrated in the inset of Fig. \ref{FS}(c), another hole FS ($\Gamma'$)  appears around $(\pi, \pi)$ only when holes are doped and/or a pnictogen height($h_{\rm As}$) from the Fe plane is relatively high as in Ba$_{0.6}$K$_{0.4}$Fe$_{2}$As$_{2}$, but it may be absent in electron-doped La1111 compounds\cite{Ikeda2,Kuroki2}.  
In fact, the $h_{\rm As}$s of both Ba$_{0.6}$K$_{0.4}$Fe$_{2}$As$_{2}$ and Nd1111($T_c= 54$ K) are approximately $\sim 1.38$\AA, being higher than that of the La1111 system($\sim 1.33$\AA).
If such evolution of FS topologies from hole-doped to electron-doped systems and band structure effect are assumed, the doping dependence of $1/T_1T$ in the normal state can be qualitatively reproduced by the theoretical calculation based on fluctuation-exchange approximation\cite{Ikeda,Ikeda2}.
In this context, the investigation of the {\it electron-doped} $Ln$1111 systems with the highest $T_c$ of more than $50$ K is highly desired to shed light on a possible correlation between $T_c$ and an evolution of FS topologies.

In summary, nuclear spin relaxation rate measurement through the $^{75}$As-NQR has revealed that the coherence effect on the Cooper pair partially recovers when the hole FS becomes considerably smaller in the heavily electron-doped La1111 compound with $T_c=5$ K.  
This is in contrast to the previous results reported in the optimally doped La1111 with $T_c=28$ K where the coherence effect was strongly suppressed. 
The marked difference is reasonably accounted for by the fact that a sign-nonconserving interband scattering process is significantly larger than the sign-conserving intraband scattering process in optimally-doped regimes, whereas the both contributions are nearly comparable in heavily overdoped regimes.
It is derived by the difference in Fermi surface topology depending on the electron doping levels. 
Our findings strongly suggest that the sign reversal of the gap functions on the different FSs, that is, the $s_\pm$-wave state is realized in La1111 compounds. 
We remark that the evolution of  $T_c$ up to 28 K in LaFeAsO systems originates from the realization of the sign reversing $s_\pm$-wave state stabilized by the interband scattering effect. 
This unconventional SC state observed in the Fe-based pnictides opens up a new paradigm for understanding a rich variety of SC phenomena. 


We would like to thank H. Ikeda and K. Kuroki for valuable comments. This work was supported by a Grant-in-Aid for Specially Promoted Research (20001004) and by the Global COE Program (Core Research and Engineering of Advanced Materials-Interdisciplinary Education Center for Materials Science) from the MEXT, Japan.




\begin{thebibliography}{99} 

\bibitem{BCS} J. Bardeen, L. N. Cooper, and J. R. Schrieffer: Phys. Rev. {\bf 108} (1957) 1175.
\bibitem{Tinkham} M. Tinkham: Introduction to Superconductivity (McGraw-Hill, New York, 1975).
\bibitem{HS} L. C. Hebel and C. P. Slichter: Phys. Rev. {\bf 113} (1959) 1504.
\bibitem{Kamihara2008} Y. Kamihara, T. Watanabe, M. Hirano, and H. Hosono: J. Am. Chem. Soc. {\bf 130} (2008) 3296.
\bibitem{Singh} D. J. Singh and M. -H. Du: Phys. Rev. Lett. {\bf 100} (2008) 237003.
\bibitem{Cruz} C. de la Cruz, Q. Huang, J. W. Lynn, J. Y. Li, W. Ratcliff, II, J. L. Zarestky, H. A. Mook, G. F. Chen, J. L. Luo, N. L. Wang, and P. C. Dai: Nature (London) {\bf 453} (2008) 899. 
\bibitem{Mazin} I. I. Mazin, D. J. Singh, M. D. Johannes, and M. H. Du: Phys. Rev. Lett. {\bf 101} (2008) 057003.
\bibitem{Kuroki} K. Kuroki, S. Onari, R. Arita, H. Usui, Y. Tanaka, H. Kontani, and H. Aoki: Phys. Rev. Lett. {\bf 101} (2008) 087004. 
\bibitem{Ding} H. Ding, P. Richard, K. Nakayama, K. Sugawara, T. Arakane, Y. Sekiba, A. Takayama, S. Souma, T. Sato, T. Takahashi, Z. Wang, X. Dai, Z. Fang, G. F. Chen, J. L. Luo, and N. L. Wang: Europhys. Lett. {\bf 83} (2008) 47001. 
\bibitem{Hashimoto} K. Hashimoto, T. Shibauchi, T. Kato, K. Ikada, R. Okazaki, H. Shishido, M. Ishikado, H. Kito, A. Iyo, H. Eisaki, S. Shamoto, and Y. Matsuda: Phys. Rev. Lett. {\bf 102} (2009) 017002.
\bibitem{Nakai} Y. Nakai, K. Ishida, Y. Kamihara, M. Hirano, and H. Hosono: J. Phys. Soc. Jpn. {\bf 77} (2008) 073701.
\bibitem{Graf} H. -J. Grafe, D. Paar, G. Lang, N. J. Curro, G. Behr, J. Werner, J. Hamann-Borrero, C. Hess, N. Leps, R. Klingeler, and B. B\"uchner: Phys. Rev. Lett. {\bf 101} (2008) 047003.
\bibitem{MukudaNQR} H. Mukuda, N. Terasaki, H. Kinouchi, M. Yashima, Y. Kitaoka, S. Suzuki, S. Miyasaka, S. Tajima, K. Miyazawa, P. M. Shirage, H. Kito, H. Eisaki, and A. Iyo: J. Phys. Soc. Jpn. {\bf 77} (2008) 093704. 
\bibitem{Kawasaki} S. Kawasaki, K. Shimada, G. F. Chen, J. L. Luo, N. L. Wang, and G. -q. Zheng: Phys. Rev. B {\bf 80} (2008) 060503. 
\bibitem{NMRtheory1} D. Parker, O. V. Dolgov, M. M. Korshunov, A. A. Golubov, and I. I. Mazin: Phys. Rev. B {\bf 78} (2008) 134524.
\bibitem{NMRtheory1a} A.V. Chubukov, D. V. Efremov, and I. Eremin: Phys. Rev. B {\bf 78} (2008) 134512. 
\bibitem{NMRtheory1b} Y. Bang and H. Y. Choi: Phys. Rev. B {\bf 78}, (2008) 134523. 
\bibitem{NMRtheory1c} M. M. Parish, J. Hu, and B. A. Bernevig: Phys. Rev. B {\bf 78} (2008) 144514. 
\bibitem{NMRtheory2} Y. Nagai, N. Hayashi, N. Nakai, H. Nakamura, M. Okumura, and M. Machida: New J. Phys. {\bf 10} (2008) 103026.
\bibitem{Terasaki} N. Terasaki, H. Mukuda, M. Yashima, Y. Kitaoka, K. Miyazawa, P. M. Shirage, H. Kito, H. Eisaki, and A. Iyo: J. Phys. Soc. Jpn. {\bf 78} (2009) 013701.
\bibitem{MukudaFe2} H. Mukuda, N. Terasaki, N. Tamura, H. Kinouchi, M. Yashima, Y. Kitaoka, K. Miyazawa, P. M. Shirage, S. Suzuki, S. Miyasaka, S. Tajima, H. Kito, H. Eisaki, and A. Iyo: J. Phys. Soc. Jpn. {\bf 78} (2009) 084717. 
\bibitem{MukudaPhysC} H. Mukuda, N. Terasaki, M. Yashima, H. Nishimura, Y. Kitaoka, and A. Iyo: Physica C {\bf 469} (2009) 559. 
\bibitem{SKitagawa} S. Kitagawa, Y. Nakai, T. Iye, K. Ishida, Y. Kamihara, M. Hirano, and H. Hosono: Physica C (2009), {\it in press}[DOI:10.1016/j.physc.2009.10.106].
\bibitem{Lang} G. Lang, H. -J. Grafe, D. Paar, F. Hammerath, K. Manthey, G. Behr, J. Werner, and B. B\"uchner: Phys. Rev. Lett. {\bf 104} (2010) 097001.
\bibitem{Ikeda} H. Ikeda: J. Phys. Soc. Jpn. {\bf 77} (2008) 123707.
\bibitem{Yashima} M. Yashima, H. Nishimura, H. Mukuda, Y. Kitaoka, K. Miyazawa, P. M. Shirage, K. Kihou, H. Kito, H. Eisaki, and A. Iyo: J. Phys. Soc. Jpn. {\bf 78} (2009) 103702.
\bibitem{Yamashita2} H. Yamashita, H. Mukuda, M. Yashima, S. Furukawa, Y. Kitaoka, K. Miyazawa, P. M. Shirage, H. Eisaki, and A. Iyo: J. Phys. Soc. Jpn. {\bf 79} (2010) No.10, {\it in press}. 
\bibitem{Sekiba} Y. Sekiba, T. Sato, K. Nakayama, K. Terashima, P. Richard, J. H. Bowen, H. Ding, Y. -M. Xu, L. J. Li, G. H. Cao, Z. -A. Xu, and T. Takahashi: New J. Phys. {\bf 11} (2009) 025020.
\bibitem{Hanaguri} T. Hanaguri, S. Niitaka, K. Kuroki, and H. Takagi: Science {\bf 328} (2010) 5977.
\bibitem{Kuroki2} K. Kuroki, H. Usui, S. Onari, R. Arita, and H. Aoki: Phys. Rev. B {\bf 79} (2009) 224511.
\bibitem{Ikeda2} H. Ikeda, R. Arita, and J. Kune$\check{\rm s}$: Phys. Rev. B {\bf 82} (2010) 024508. 



\end{thebibliography}
\end{document}